\documentclass[conference]{IEEEtran}
\pdfoutput=1
\IEEEoverridecommandlockouts
\usepackage{amsmath,amssymb,amsfonts}

\usepackage{algorithm}
\usepackage{algorithmicx}
\usepackage{algpseudocode}
\usepackage{graphicx}
\usepackage{textcomp}
\usepackage{xcolor}
\usepackage{epstopdf}
\usepackage[numbers]{natbib}
\usepackage{subfig}
\usepackage{multicol}
\usepackage{lipsum}
\usepackage{enumerate}

\newtheorem{lemma}{Lemma}
\newtheorem{theorem}{Theorem}

\newtheorem{definition}{Definition}

\def\BibTeX{{\rm B\kern-.05em{\sc i\kern-.025em b}\kern-.08em
    T\kern-.1667em\lower.7ex\hbox{E}\kern-.125emX}}
\begin{document}

\title{Efficient Two-Dimensional Self-Stabilizing Byzantine Clock Synchronization in WALDEN
\thanks{This work has been accepted by ICPADS2021.\\
© 2022 IEEE.  Personal use of this material is permitted.  Permission from IEEE must be obtained for all other uses, in any current or future media, including reprinting/republishing this material for advertising or promotional purposes, creating new collective works, for resale or redistribution to servers or lists, or reuse of any copyrighted component of this work in other works.}
}

\author{\IEEEauthorblockN{1\textsuperscript{st} Shaolin Yu}
\IEEEauthorblockA{\textit{Tsinghua University}\\
Beijing, China \\
ysl8088@163.com}
\and
\IEEEauthorblockN{2\textsuperscript{nd} Jihong Zhu}
\IEEEauthorblockA{\textit{Tsinghua University}\\
Beijing, China \\
jhzhu@tsinghua.edu.cn}
\and
\IEEEauthorblockN{3\textsuperscript{rd} Jiali Yang}
\IEEEauthorblockA{\textit{Tsinghua University}\\
Beijing, China \\
yangjiali-0411@163.com}
}

\maketitle

\begin{abstract}
For tolerating Byzantine faults of both the terminal and communication components in self-stabilizing clock synchronization, the two-dimensional self-stabilizing Byzantine-fault-tolerant clock synchronization problem is investigated and solved.
By utilizing the time-triggered (TT) stage provided in the underlying networks as TT communication windows, the approximate agreement, hopping procedure, and randomized grandmasters are integrated into the overall solution.
It is shown that with partitioning the communication components into $3$ arbitrarily connected subnetworks, efficient synchronization can be achieved with one such subnetwork and less than $1/3$ terminal components being Byzantine.
Meanwhile, the desired stabilization can be reached for the specific networks in one or several seconds with high probabilities.
This helps in developing various distributed hard-real-time systems with stringent time, resources, and safety requirements.
\end{abstract}


\section{Introduction}
Reliable synchronization is fundamental in constructing reliable distributed hard-real-time systems.
For example, in constructing distributed integrated modular avionics (DIMA) systems under the popular time-triggered (TT) architecture \citep{RN498,as6802}, the designers should show sufficiently high assumption coverage \citep{Powell1992assumption} of the provided fault-tolerance schemes.
In this background, to avoid common-mode failures\citep{Lala1994Architectural} as far as possible, a reliable global synchronization scheme is the basis for developing high-layer fault-tolerant systems.
In providing such synchronization schemes for safety-critical hard-real-time systems, self-stabilizing Byzantine-fault-tolerant clock synchronization (SS-BFT-CS) is a promising trend.

Here, the terminology \emph{SS-BFT-CS} is at the intersection of three widely investigated concepts in the field of distributed systems.
Firstly, clock synchronization (CS) is the basis for coordinating the distributed (and often redundant) components of the hard-real-time systems.
For such a CS scheme to be reliable, the provided CS algorithms should tolerate various kinds of faults of the components.
As it is not easy to show that the failure modes of the faulty components can always be restricted with sufficiently high probabilities, it is safe to assume that the faulty components can fail arbitrarily, i.e., to allow them being \emph{Byzantine}.
Under this assumption, the Byzantine-fault-tolerant clock synchronization (BFT-CS) problem is intensively studied in both the theoretical and the industrial realms.
Further, in considering that the system might experience some unforeseen transient disturbances, it is also hard to show there cannot be $1/3$ \citep{RN3905} or more Byzantine components in the system under all working conditions.
In this context, self-stabilization can take a great role in bringing the system from arbitrary initial states to the desired stabilized states.
Following this trend, the widely adopted TT solutions, including the Time-Triggered Protocol (TTP) \citep{RN498} and the Time-Triggered Ethernet (TTEthernet) \citep{as6802}, are all built upon some practical SS-BFT-CS schemes.

However, although some industrial CS solutions are integrated with both Byzantine-fault-tolerance and self-stabilization, only Byzantine faults of the terminal components can be tolerated in the provided CS algorithms.
Meanwhile, the failure-modes of some communication components are still gravely restricted.
For example, in TTP, as the number of the central guardians employed in the redundant bus network is typically restricted, these guardians are not allowed to be Byzantine.
In TTEthernet, as the network connectivity of typical switched Ethernet is still low, the TT-switches are not allowed to be Byzantine.
In this situation, to show sufficiently high assumption coverage of the CS schemes, these core communication components should be designed, manufactured, and verified with great effort.
This would restrict the applications of these CS schemes to a broader range.

\begin{figure}[htbp]
\centerline{\includegraphics[width=3.0in]{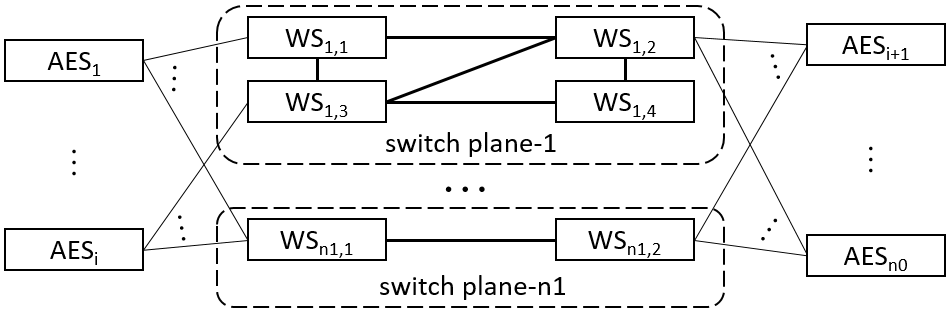}}
\caption{ A WALDEN network with multiple switch planes.}
\label{fig:WALDEN Network}
\end{figure}

In this paper, to provide high-reliable CS, we investigate the SS-BFT-CS problem in the presence of Byzantine faults of both the terminal components and the communication components.
For a concrete study, we would extend our former work, the basic WALDEN network \cite{YuCOTS2021}, to the advanced ones with multiple-Ethernet-interface terminal components.
Typically, as the network connectivity of switched-Ethernet is not sufficiently high, it is often impossible to locally prevent the propagation of the Byzantine faults generated in the communication components.
In preventing the propagation of such faults, we will partition the WALDEN switches (the WS nodes) into several isolated subnetworks, as is shown in Fig.~\ref{fig:WALDEN Network}.

Here, each of the isolated subnetworks of the WS nodes can be viewed as in an independent switch plane.
In each such switch plane, we only require that the WS nodes are connected with an arbitrary topology.
For tolerating one or more such subnetworks being fail-arbitrarily, we require that each terminal component is connected to several such subnetworks.
In the words of \cite{YuCOTS2021}, such a terminal component is called an advanced end-system (AES).
In the whole picture, the isolated subnetworks can also be interconnected for upper-layer applications, as long as the synchronization signals and messages transmitted in these interconnections (sent by faulty nodes) would never be received by the nonfaulty nodes.

We can see, this networking scheme can provide a good tradeoff between performance and fault-tolerance.
For performance, firstly, as there can be several switches in each subnetwork, sufficient communication bandwidth and connections can be provided for high-layer communications.
Secondly, by designing and optimizing TDMA schedules in each switch plane with arbitrary network topologies, the overall bandwidth resources can be better allocated and utilized.
Meanwhile, for fault-tolerance, as the TT communication can be realized in the WALDEN switches, the faults of terminal nodes can be isolated with static TDMA schedules.
Also, as the faulty synchronization signals generated in the faulty switches cannot propagate between different subnetworks, each switch plane can be viewed as a fault-containment region (FCR\citep{RN896}, or FTU\citep{RN498}).
With this, the redundant switch planes are orthogonal to the redundant terminal nodes in the view of fault-tolerance.

So, the main problem is to provide an efficient SS-BFT-CS solution for advanced WALDEN networks with the two-dimensional Byzantine resilience, in which $f_0$ Byzantine terminal nodes and $f_1$ Byzantine switch planes can be tolerated at the same time.
For this, the rest of this paper is constructed as follows.
The related work and basic definitions are respectively given in Section~\ref{sec:related} and Section~\ref{sec:model}.
In Section~\ref{sec:strategies}, the synchronization strategies are presented intuitively.
With this, the SS-BFT-CS algorithm is provided, analyzed, and discussed in Section~\ref{sec:algo}.
Lastly, we conclude the paper in Section~\ref{sec:con}.

\section{Related work}
\label{sec:related}
In the literature, several SS-BFT-CS solutions have been provided in both the deterministic and probabilistic approaches, but most of them can only tolerate Byzantine faults of the terminal nodes.
In deterministic approaches, linear stabilization time solutions \citep{Daliot2006Linear,DolevPulseBoundedDelay2007} are provided by employing some self-stabilizing Byzantine agreement protocols as building blocks \cite{Daliot2006Agreement}.
However, as is discussed in \cite{Dolev2014PulseGeneration}, these BA-based SS-BFT-CS solutions are often with poor performance in considering the required message and computation complexities.
In \cite{Lenzen2019AlmostasEasyas}, by iteratively constructing the higher-layer SS-BFT-CS algorithms with the lower-layer resynchronization algorithms and consensus routines as building blocks, it is shown that the BA-based SS-BFT-CS solutions can be almost as easy as consensus.
However, the message complexity and network connectivity required in the consensus routines still gravely restrict the wide application of these solutions in real-world sparsely connected networks.

In probabilistic approaches, \cite{DolevWelchSelf2004} first provides a randomized SS-BFT-CS solution with intuitive hopping procedures with expected hyper-exponential stabilization time.
Later in \cite{Dolev2014PulseGeneration}, the expectation of the stabilization time is lowered to $O(n)$, where $n$ is the number of the nodes in the system.
Another advantage of \cite{Dolev2014PulseGeneration} is the very little requirement on computation resources.
However, to apply \cite{Dolev2014PulseGeneration}, the communication network is required to be fully connected.
Meanwhile, the frequency of the synchronization signals is high, since the solution is mainly for synchronizing some tiny components deployed in VLSI circuits within the range of several centimeters.

In the industrial realm, TTEthernet \citep{as6802} provides standard and high-reliable SS-BFT-CS solutions.
However, Byzantine faults of the TT-switches should be locally guarded and filtered with special hardware schemes like the monitor-pairs \citep{as6802}.
In \citep{YuCOTS2021}, a self-stabilizing synchronization solution is provided by employing COTS Ethernet components as building blocks, but only Byzantine terminal nodes can be tolerated.
Further, for tolerating Byzantine communication components, although we may provide some multi-path transmission schemes with the multiple switch planes, the stabilization time of SS-BFT-CS would be restricted by the number of the Byzantine terminal nodes.
With our limited knowledge, no efficient SS-BFT-CS solution tolerating Byzantine faults of both the terminal and communication components exists with the stabilization time being independent of the number of the Byzantine terminal nodes.

\section{System model and the problem}
\label{sec:model}
\subsection{Basic assumptions}
Following \cite{YuCOTS2021}, the WALDEN network $G(V,E)$ is comprised of three kinds of nodes: the master end-system (MES) nodes $V_A$, the WALDEN switch (WS) nodes $V_B$, and the client end-system (CES) nodes $V_C$.
In discussing the core SS-BFT-CS problem, the CES nodes are ignored.
In considering fault-tolerance, the faulty nodes in $V$ are denoted as $F$, with which the nonfaulty nodes are denoted as $Q=V\setminus F$.
For convenience, we also denote $Q_x=Q\cap V_x$ and $F_x=F\cap V_x$ for $x\in \{A,B\}$.
To our goal, all nodes in $F$ are allowed to fail arbitrarily.
Moreover, in considering self-stabilization, every node in $Q$ runs the given algorithms since some instant $t_0$ with an arbitrary initial state.

Since $t_0$, in running the synchronization algorithms given in this paper, we make the following assumptions.
Firstly, every node $i\in Q_A$ can send and receive standard Ethernet messages (the standard messages).
Every node $j\in Q_B$ can send, receive, and deliver standard messages.
Besides, every node $j\in Q_B$ can also propagate pulse-like synchronization signals (the \emph{SIGs}) to other connected WS nodes in the same switch plane.
As is in \cite{YuCOTS2021}, the propagation of the \emph{SIGs} has a higher preemptive priority than the propagation (including sending, delivering, and receiving) of the standard messages.
For simplicity, the preempted standard messages would be discarded in the nonfaulty WS nodes.
In each nonfaulty subnetwork (including the terminal nodes), if a standard message is not preempted, it can be propagated (including sent, delivered, received, and processed) within $d_{max}$ time.

Secondly, for approximately measure the \emph{time} (here the \emph{time} can be taken as the Newtonian time for simplicity), every node $i\in Q$ is equipped with a hardware clock $H_i$ that can count the periodically generated ticking events with a tick-counter.
For accuracy, the ticking cycles of $H_i$ are always bounded in $[(1-\rho)T_H,(1+\rho)T_H]$, where $T_H$ is the nominal ticking cycle, and $\rho$ is the maximal drift-rate of the hardware clocks.
At any given instant $t$, the current value of the tick-counter in $H_i$ is denoted as $H_i(t)$.
In real-world systems, we assume $H_i(t)$ can only take values in $[[\tau_{max}]]$, where $[[x]]=\{0,1,\dots,x-1\}$ is the set of the first $x$ non-negative integers and $\tau_{max}$ is sufficiently large.
In considering self-stabilization, $H_i(t_0)$ can be arbitrarily valued in $[[\tau_{max}]]$ in every $i\in Q$.
Denoting $t_1$ and $t_2$ as two adjacent ticking instants of $H_i$ with $t_2>t_1>t_0$, we always have $H_i(t_2)=(H_i(t_1)+1)\bmod \tau_{max}$.
As $H_i$ is often used as an unadjustable timing source (for implementing various timers), we define the adjustable local clock $C_i$ in each node $i\in Q$.
By maintaining a local variable $\mathtt{offset}_i$ in node $i$, the value of $C_i$ at $t$ can be represented as $C_i(t)=(H_i(t)+\mathtt{offset}_i(t))\bmod \tau_{max}$, with which the accuracy of $H_i$ is also shared in $C_i$.
For convenience, as the clocks all take values in $[[\tau_{max}]]$, we denote $\tau_1\oplus\tau_2=(\tau_1+\tau_2)\bmod \tau_{max}$, $\tau_1\ominus\tau_2=(\tau_1-\tau_2)\bmod \tau_{max}$, and $\mathring{d}(\tau_1,\tau_2)=\min\{\tau_1\ominus\tau_2,\tau_2\ominus\tau_1\}$.

\subsection{The advanced WALDEN network and the problem}

As is shown in Fig.~\ref{fig:WALDEN Network}, we assume that the WALDEN network $G=(V_A \cup V_B, E_A \cup E_B)$ is composed of $n_0=|V_A|$ MES nodes and $n_1$ switch subnetworks.
The $s$th ($s\in S=\{1,2,\dots, n_1\}$) switch subnetwork can be represented as a connected graph $G_B^{(s)}=G(V_B^{(s)}, E_B^{(s)})$ where $V_B^{(s)}$ and $E_B^{(s)}$ are respectively the WS nodes and edges on the $s$th switch plane.
Then, every $G_B^{(s)}$ is connected to the MES nodes $V_A$ with the edges $E_A^{(s)}$.
For clarity here, we have $\bigcup_{s=1}^{n_1}V_B^{(s)}=V_B$, $\bigcup_{s=1}^{n_1}E_B^{(s)}=E_B$, $\bigcup_{s=1}^{n_1}E_A^{(s)}=E_A$, $V_B^{(s_1)}\cap V_B^{(s_2)}=\emptyset$, $E_B^{(s_1)}\cap E_B^{(s_2)}=\emptyset$, and $E_A^{(s_1)}\cap E_A^{(s_2)}=\emptyset$ for all $s,s_1,s_2\in S$.
With this, each subnetwork $G(V_A \cup V_B^{(s)}, E_A^{(s)}\cup E_B^{(s)})$ can be viewed as a basic WALDEN network in the $s$th network plane.

Typically, as each switch subnetwork $G_B^{(s)}$ has a relatively low network connectivity, each $G_B^{(s)}$ is viewed as an FCR, i.e., $G_B^{(s)}$ (and the corresponding $s$th switch plane) is regarded as \emph{nonfaulty} iff all nodes and edges in $G_B^{(s)}$ are \emph{nonfaulty}.
Meanwhile, each advanced MES node $i\in V_A$ is viewed as an FCR.
To our goal, only the nodes in the nonfaulty FCRs are required to be synchronized.
For convenience, we also use $V_P=S$ to denote the set of all switch planes and use $Q_P\subseteq V_P$ to denote the nonfaulty switch planes.
And the set of the considered nonfaulty nodes is denoted as $Q_{AP}=Q_A\cup \bigcup_{s\in Q_P}V_B^{(s)}$.
With this, we say the system $\mathcal{S}$ is $(\epsilon_0,\varrho,\Delta)$-\emph{synchronized} in $[t_{1},t_{2}]$ iff
\begin{eqnarray}
\label{eq:time_synchronization_precision2} \mathring{d}(C_i(t),C_j(t))\leqslant \epsilon_0 \\
\label{eq:time_synchronization_accuracy2} |(C_i(t')\ominus C_i(t))-(t'-t)|\leqslant \varrho (t'-t) +\epsilon_0
\end{eqnarray}
hold for all $i,j\in Q_{AP}$ and all $t',t\in [t_{1},t_{2}]$ with $0\leqslant t'-t\leqslant\Delta$.
For simplicity, we can set $\Delta$ as approximately the synchronization cycle and thus the core problem is to make $\mathcal{S}$ being $(\epsilon_0,\rho)$-\emph{synchronized} in $[t_{1}+\infty)$.
Concretely, in the two dimensional SS-BFT-CS problem, given that all nodes in $Q$ are locally \emph{correct} \citep{YuCOTS2021,DolevPulseBoundedDelay2007} since $t_c$, $\mathcal{S}$ should be $(\epsilon_0,\rho)$-\emph{synchronized} in $[t,+\infty)$ with some $t\leqslant t_c+\Delta_\mathtt{stb}$, where the stabilization time $\Delta_\mathtt{stb}$ is expected to be sufficiently small.

At the same time, to prevent $\mathcal{S}$ from being $(\epsilon_0,\rho)$-synchronized, the adversary knows all the provided algorithms and can arbitrarily choose $f_0$ MES nodes and an arbitrary number of WS nodes in $f_1$ switch planes being Byzantine.
Also, the adversary can arbitrarily set the state of $\mathcal{S}$ at $t_0$.
Since $t_0$, all chosen Byzantine nodes are under the full control of the adversary who knows everything about the executions of $\mathcal{S}$.
Moreover, the adversary can arbitrarily choose the delays and clock drifts in their bounded ranges in the executions of $\mathcal{S}$.
Now to play the game with such a static adversary, we assume $n_0>3f_0$ and $n_1>2f_1$.

\section{Strategies}
\label{sec:strategies}
As is shown in Fig.~\ref{fig:WALDEN Network}, every advanced MES node in the advanced WALDEN network $G=(V_A \cup V_B, E_A \cup E_B)$ is connected to $n_1$ switch subnetworks to tolerate up-to $f_1$ faulty switch subnetworks.
This is the main difference from the basic WALDEN network proposed in \cite{YuCOTS2021}.
Now, in solving the two-dimensional SS-BFT-CS problem with $G$, although the advanced MES nodes now gain more resources to tolerate the faults generated in the WS nodes, the faulty MES nodes can gain \emph{more} power than that of the nonfaulty MES nodes, as the adversary can know more than the nonfaulty nodes and can collusively control the delays, clock drifts, and the Byzantine nodes.
In such a situation, we take the following strategies.
\subsection{The decoupled synchronization strategies}
Firstly, to restrict the power of faulty MES nodes, every node $i\in Q_A$ is allowed to send standard messages but not allowed to send \emph{SIG}.
Secondly, to be decoupled with the basic synchronization strategies and algorithms provided in \cite{YuCOTS2021}, every advanced MES node would act as $n_1$ independent CES nodes that parallel run the basic self-stabilizing synchronization algorithm (referred to as the SSS algorithm) provided in \cite{YuCOTS2021}.
Namely, for each switch plane $p\in V_P$, every node $i\in V_A$ can be viewed as a virtual CES node (denoted as $\mathtt{CES}(i,p)$) in running the SSS algorithm.
Thirdly, as each switch plane $p\in V_P$ is viewed as a single FCR, we can choose a specific WS node $s\in V_B^{(p)}$, denoted as $\mathtt{MWS}(p)$, to be the master-WS (MWS) node of $p$ and only allow the MWS node $s$ to generate new \emph{SIG} signals in $p$.
And when $p$ is nonfaulty, all nodes in $V_B^{(p)}$ can propagate the \emph{SIGs}.
With this, for every $p\in Q_P$, the basic synchronization round \citep{YuCOTS2021} can be periodically initiated in $p$ by the \emph{SIG} generated in $\mathtt{MWS}(p)$.

Now to synchronize the $C$ clocks of all nodes in $Q_{AP}$, we want to use the $C$ clocks of the MWS nodes $V_S=\{\mathtt{MWS}(p)\mid p\in V_P\}$ as the master $C$ clocks to synchronize the $C$ clocks (referred to as the client $C$ clocks) of all the other nodes in $Q_{AP}$.
Namely, denoting all MWS nodes in the nonfaulty switch planes as $Q_S=\{\mathtt{MWS}(p)\mid p\in Q_P\}$, if the $C$ clocks of all nodes in $Q_S$ are synchronized, the $C$ clocks of all WS nodes in $Q_{AP}$ can be trivially synchronized as the client $C$ clocks.
Also, the $C$ clock of every node $i\in Q_A$ can also be synchronized by taking the median of the $n_1$ client $C$ clocks of the parallel virtual CES nodes $\mathtt{CES}(i,p)$ for all $p\in V_P$.

So, the core problem is to synchronize the $C$ clocks of the nodes in $Q_S$.
For this, firstly, as the basic synchronization round (\emph{round} for short) can be periodically initiated by the MWS node $s=\mathtt{MWS}(p)$ in every nonfaulty switch plane $p$, we want to use the TT stage in every round of $p$ to record the $C$ clock of $\mathtt{MWS}(p)$ in every node $i\in Q_A$ and then relay this information to other nodes in $Q_S$.
As is introduced in \cite{YuCOTS2021}, the TT stage can be used to provide statically scheduled TT communication in each nonfaulty switch plane.
Namely, by running the SSS algorithms in each plane, the ticks in the TT stage can be partitioned into TT-slots with fixed TT-schedules.

\begin{figure}[htbp]
\centerline{\includegraphics[width=3.2in]{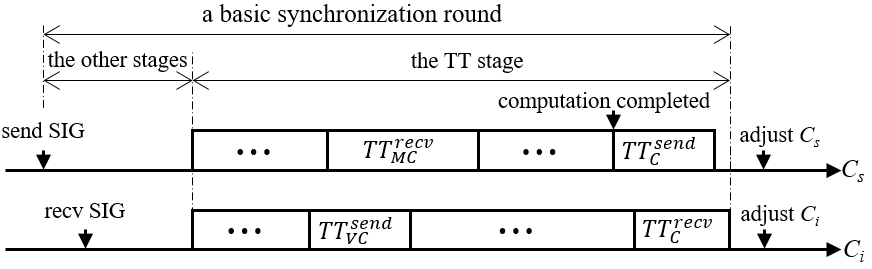}}
\caption{A basic synchronization round in a switch plane.}
\label{fig:TTstages}
\end{figure}

By utilizing the TT stages to exchange the $C$ clocks and other information, the upper-layer synchronization strategies can be well-decoupled with the basic ones provided in \cite{YuCOTS2021}.
As is shown in Fig.~\ref{fig:TTstages}, denoting the TT-slot of sending the clock information in the MWS node as $\mathtt{TT}_{C}^{send}$, $s$ can send a TT-message containing the clock information of $s$ to all nodes in $V_A$ during $\mathtt{TT}_{C}^{send}$ of $s$.
Then, during the same round, every node $i\in Q_A$ can receive and record the TT-message of $s$ during the TT-slot $\mathtt{TT}_{C}^{recv}$ of $i$.
Thus, as $s$ can initiate a new round within a bounded time, some historical messages of every node in $Q_S$ can be recorded in every node $i\in Q_A$ with timestamps.
So, after a fixed duration, when $s$ initiates a new round, every node $i\in Q_A$ can send an $n_1$ dimensional vector of the records of the MWS nodes to $s$ during the TT-slot $\mathtt{TT}_{VC}^{send}$ of $i$.
Then, during the same round, $s$ can receive an $n_1\times n_0$ matrix of the records during the TT-slot $\mathtt{TT}_{MC}^{recv}$ of $s$.
With this, $s$ can compute a new value of its $C$ clock and broadcast it in the TT-message sent during $\mathtt{TT}_{C}^{send}$.
Then, $s$ can adjust its $C$ clock at the end of every TT stage or later.

Denoting the scheduled beginning and end ticks of a TT-slot $\mathtt{TT}_x^y$ as respectively $\mathtt{begin}(\mathtt{TT}_x^y)$ and $\mathtt{end}(\mathtt{TT}_x^y)$, the scheduled TT-slots between $\mathtt{begin}(\mathtt{TT}_{MC}^{recv})$ and $\mathtt{end}(\mathtt{TT}_{C}^{send})$ are referred to as the TT-exchanging window.
With the TT-exchanging windows, every $s\in Q_S$ can distribute its newly computed $C$ clock value to $V_A$ and collect such values relayed by $V_A$ in an $n_1\times n_0$ matrix $\mathbf{M}^{(s)}$.
Also, every $i\in Q_A$ can estimate and send the current $C$ clock values of every node in $V_S$ and then every $s$ can collect such values in another $n_1\times n_0$ matrix $\mathbf{C}^{(s)}$.
With this, the main problem is to compute the new value of $C_s$ in every $s\in Q_S$ with the received TT-messages in every TT-exchanging window.
For SS-BFT-CS, two subproblems should be considered.
One is to establish the initial synchronization with the arbitrary state of the system.
The other one is to maintain the synchronized state of the system.

\subsection{The randomized grandmasters}

To establish the initial synchronization with an arbitrary system state since $t_0$, firstly, as the rounds in different switch planes are initially asynchronous, the items in $\mathbf{M}^{(s)}$ and $\mathbf{C}^{(s)}$ may be outdated.
Also, as the $C$ clocks of the MWS nodes may be symmetrically distributed in $[[\tau_{max}]]$ at $t_0$, some asymmetric operations should be taken to break this possible symmetry.
On the one hand, in breaking the symmetry, we want to take the $C$ clock of some specific MWS node to coordinate the $C$ clocks of other nodes, just like the \emph{grandmaster} employed in PTP \citep{ieee1588v2}.
On the other hand, as the adversary knows everything about the provided algorithms and the executions, this \emph{grandmaster} node cannot be fixed in the algorithms.
In this situation, one possible way is to elect a unique \emph{grandmaster}.
However, as the elected \emph{grandmaster} might be in the faulty switch plane and thus be deliberately controlled by the adversary, the stabilization of the system cannot be reached in a deterministic way.
Meanwhile, the temporally \emph{synchronized} state of the system cannot be deterministically maintained if the elected \emph{grandmaster} is controlled by the adversary.
Also, the election of the \emph{grandmaster} would be hard, if it is not impossible, in the network $G$ where the communication rounds in different switch planes are asynchronous and nearly one-half of these switch planes can fail arbitrarily.

Instead, the \emph{grandmaster} is randomly chosen.
Concretely, to be a \emph{grandmaster}, every node $s\in Q_S$ would toss a biased coin $b_\mathtt{coin}^{(s)}\in\{0,1\}$ before $s$ computing its new $C$ clock value during every round.
And if $b_\mathtt{coin}^{(s)}=1$, $s$ would regard itself as the \emph{grandmaster} during the current and the next several rounds.
Otherwise, if $b_\mathtt{coin}^{(s)}=0$ during the current and several previous rounds, $s$ would not regard itself as the \emph{grandmaster}.

Obviously, in this way, the number of the \emph{grandmasters} in $Q_S$ is random.
Nevertheless, there is a probability that a unique \emph{grandmaster} exists in $Q_S$ during a sufficiently long duration.
For a specific example, if we partition the WS nodes in the communication system into $3$ subsets and employ the nodes in every such subset to form a switch subnetwork in $G$, there would be $n_1=3$ MWS nodes, in which one MWS can be in a faulty switch subnetwork.
In this case, only two MWS nodes need to be synchronized in establishing the initial synchronization.
Namely, if only two nodes in $Q_S$ are synchronized, the required minimal synchrony of the system is established.

\subsection{The filters and the guarded conditions}

Meanwhile, to deterministically maintain the synchronized state of the system, when a node $s\in Q_S$ cannot observe any evidence of an unsynchronized state, $s$ should try to maintain the \emph{possibly} synchronized state of the system.
So, whenever $s$ finds that the collected records can be generated by $n_1-f_1$ or more synchronized nodes in $Q_S$, $s$ would assume that the system is synchronized and try to participate in some synchronization-maintaining routine.

However, when $s$ observes that the matrix $\mathbf{C}^{(s)}$ might be generated in the synchronized system, as $f_1$ rows in $\mathbf{C}^{(s)}$ can be from the faulty switch planes, there might be only $n_1-2f_1$ nodes in $Q_S$ are synchronized.
As we only require $n_1>2f_1$, there might be only one row in $\mathbf{C}^{(s)}$ that can be used as reference.
In this situation, two different nodes $s,s'\in Q_S$ can get very different references from $\mathbf{C}^{(s)}$ and $\mathbf{C}^{(s')}$ in the presence of the $f_1$ Byzantine switch planes.
Nevertheless, as there are $n_0>3f_0$ MES nodes, we can leverage the nonfaulty nodes in $Q_A$ to filter the rows of $\mathbf{C}^{(s)}$ for every $p\in V_P$.
Concretely, with the recorded $C$ clocks of the MWS nodes, every MES node $i$ can check the basic accuracy condition $a_{p,i}^{curr}$ for every switch plane $p\in V_P$.
Denoting the instants of receiving the current and the previous records from $p\in V_P$ in $i$ as $t_{curr}$ and $t_{pre}$, $a_{p,i}^{curr}$ can be computed as
\begin{eqnarray}
\label{eq:accuracy_condition}\mathring{d}(m_{p,i}(t_{curr}), m_{p,i}(t_{pre})\oplus T)\leqslant 2\epsilon_0 \land \mathring{d}(H_i(t_{curr}),\nonumber\\
 H_i(t_{pre})\oplus T)\leqslant (2\epsilon_0+2\rho T+d_{max})/(1-\rho)^2
\end{eqnarray}
where $m_{p,i}(t)$ is the received value from $p$ in node $i$ at $t$.

Then, $i$ can maintain an accuracy-counter $a_{p,i}$ to count the rounds in which $a_{p,i}^{curr}$ is always true.
To our aim, the value of $a_{p,i}$ can be upper-bounded by $a_0=3$.
And once $a_{p,i}^{curr}$ is not true in the current round, $a_{p,i}$ would be reset to $0$.
With this, besides sending the estimated values of the $C$ clocks of the MWS nodes, every node $i\in Q_A$ is also scheduled to send $a_{p,i}$ during the TT-slot $\mathtt{TT}_{VC}^{send}$ of $i$.
Then, in every node $s\in Q_S$, by collecting these $a_{p,i}$ into the $n_1\times n_0$ matrix $\mathbf{A}^{(s)}$, if the $p$th row of $\mathbf{A}^{(s)}$ contains $n_0-f_0$ or more $a_0$, $p$ would pass the accuracy filter and would be put in $P_\mathtt{acc}^{(s)}$.
Otherwise, $p$ would be filtered out and would not be put in $P_\mathtt{acc}^{(s)}$.

Besides, when the system is synchronized, as the recorded messages of every $p\in Q_P$ in all nonfaulty MES nodes would be the same value sent by $\mathtt{MWS}(p)$, there would be at least $n_0-f_0$ items in the $p$th row of $\mathbf{M}^{(s)}$ being with the same value.
So we can define
\begin{eqnarray}
\label{eq:majority_filter}P_\mathtt{maj}^{(s)}=\{p\in V_P\mid \exists V'\subseteq V_A: |V'|\geqslant n_0-f_0\land \nonumber\\
\forall i\in V': m_{p,i}^{(s)}=m_p^{(s)}\}
\end{eqnarray}
where $m_p^{(s)}=\mathtt{med}(m_{p,1}^{(s)},\dots,m_{p,n_0}^{(s)})$ is the majority value in the $p$th row of $\mathbf{M}^{(s)}$ with $\mathtt{med}$ being the median function and $m_{p,i}^{(s)}$ being the $p$th row $i$th column element of $\mathbf{M}^{(s)}$.

So, by collecting the row numbers in $P_\mathtt{acma}^{(s)}=P_\mathtt{acc}^{(s)}\cap P_\mathtt{maj}^{(s)}$, only the $P_\mathtt{acma}^{(s)}$ rows of $\mathbf{C}^{(s)}$ need to be considered in deciding if $\mathbf{C}^{(s)}$ might be generated in the synchronized system.
With this, by configuring a sufficiently large $\epsilon_1=O(\epsilon_0+\rho T+d_{max})$, the stabilization condition $\mathtt{E}_\mathtt{stb}^{(s)}$ can be guarded in $s$ as
\begin{eqnarray}
\label{eq:condition_stb}\exists P'\subseteq P_\mathtt{acma}^{(s)},V'\subseteq V_A: |P'|\geqslant n_1-f_1\land |V'|\geqslant n_0-f_0\land \nonumber\\
\forall p_1,p_2\in P',i_1,i_2\in V':\mathring{d}(c_{p_1,i_1}^{(s)}-c_{p_2,i_2}^{(s)})\leqslant \epsilon_1
\end{eqnarray}
with $c_{p,i}^{(s)}$ being the $p$th row $i$th column element of $\mathbf{C}^{(s)}$.
However, as there can be $f_0$ faulty MES nodes, $\mathtt{E}_\mathtt{stb}$ might be inconsistently observed among the nodes in $Q_S$.
So $s$ guards another condition $\mathtt{E}_\mathtt{weak}^{(s)}$ such that if $\mathtt{E}_\mathtt{stb}^{(s')}$ is true in some $s'\in Q_S$, $\mathtt{E}_\mathtt{weak}^{(s)}$ would be true in some desired synchronization procedures.
For this, by configuring a sufficiently large $\epsilon_2=O(\epsilon_1)$, $\mathtt{E}_\mathtt{weak}^{(s)}$ can be guarded as
\begin{eqnarray}
\label{eq:condition_weak}\exists P'\subseteq V_P,V'\subseteq V_A: |P'|\geqslant n_1-f_1\land |V'|\geqslant n_0-2f_0\land \nonumber\\
\exists c_\mathtt{weak}^{(s)}\in[[\tau_{max}]]:\forall p\in P',i\in V':\mathring{d}(c_{p,i}^{(s)},c_\mathtt{weak}^{(s)})\leqslant \epsilon_2/2
\end{eqnarray}
where $c_\mathtt{weak}^{(s)}$ is referred to as a weak reference.

\subsection{Synchronization rules and fault-tolerant functions}
Now, to integrate the strategies of establishing the initial synchronization state and maintaining the synchronized state of the system, we give the following synchronization rules.

In each round of a node $s\in Q_S$, firstly, if $s$ does not regard itself as the \emph{grandmaster}, $s$ would try to be synchronized by some \emph{grandmaster}.
For this, as $s$ does not know which node is currently a \emph{grandmaster} or if there is a \emph{grandmaster}, $s$ would call a randomized fault-tolerant function (denoted as $\mathtt{RFT}$) to compute the new $C$ clock value $c_{new}^{(s)}$, providing that $\mathtt{E}_\mathtt{stb}^{(s)}$ is not true in $s$.
For $n_1=3$, we can define
\begin{eqnarray}
\label{eq:func_ftc}\mathtt{RFT}(\mathbf{C}^{(s)},c_s^{pre})=\mathtt{rd}_{p_0}(\mathtt{FTA}(\mathbf{C}^{(s)}),\mathtt{rd}(\{c_1,c_2,c_3,c_s^{pre}\}))
\end{eqnarray}
with $c_p=\mathtt{med}(c_{p,1}^{(s)},\dots,c_{p,n_0}^{(s)})$ and $c_s^{pre}$ being some previous clock value of $s$.
Here, the $\mathtt{rd}_{p_0}(x_0,x_1)$ function randomly chooses $x_0$ with the probability $p_0$ (and thus chooses $x_1$ with the probability $1-p_0$).
The $\mathtt{rd}(X)$ function randomly chooses an element of $X$ with the uniform probability.
The $\mathtt{FTA}$ function is the deterministic fault-tolerant averaging function provided in \citep{Dolev1986Approximate}, which is defined as
\begin{eqnarray}
\label{eq:func_fta}\mathtt{FTA}(\mathbf{C}^{(s)})=\mathtt{mean}(\mathtt{select}_{f_0}(\mathtt{reduce}_{f_0}(\bar{c}_1,\dots,\bar{c}_{n_0})))
\end{eqnarray}
with $\bar{c}_i=\mathtt{med}(c_{1,i}^{(s)},\dots,c_{n_1,i}^{(s)})$ and the $\mathtt{mean}$, $\mathtt{select}_{f_0}$, and $\mathtt{reduce}_{f_0}$ functions being all from \citep{Dolev1986Approximate}.
With this, if $\mathtt{E}_\mathtt{stb}^{(s)}$ is true, as $s$ should also try to maintain the synchronized state of the system, $s$ would directly call the $\mathtt{FTA}$ function (instead of the $\mathtt{RFT}$ function) to compute $c_{new}^{(s)}$.
In both cases, $s$ would send the newly computed $c_{new}^{(s)}$ during the TT-slot $\mathtt{TT}_{C}^{send}$ and use $c_{new}^{(s)}$ to adjust $C_s$ at the end of the current round.

Otherwise, if $s\in Q_S$ regards itself as the \emph{grandmaster}, $s$ would try to remain its $C$ clock being unadjusted and expect that all other nodes in $Q_S$ would choose $s$ as the \emph{grandmaster}.
However, as every $s'\in Q_S\setminus\{s\}$ should check the $\mathtt{E}_\mathtt{stb}^{(s')}$ condition to decide which function should be called in computing $c_{new}^{(s)}$, the adversary can prevent $s'$ from calling the $\mathtt{RFT}$ function by making $\mathtt{E}_\mathtt{stb}^{(s')}$ being always true in $s'$.
Also, when $s$ and $s'$ are synchronized or coarsely synchronized, $s$ should participate in the desired approximate agreement with a high probability.
So, when $b_\mathtt{coin}^{(s)}=0$ or $\mathtt{E}_\mathtt{stb}^{(s)}$ is true, $s$ would also call the $\mathtt{FTA}$ function.
Otherwise, when $\mathtt{E}_\mathtt{weak}^{(s)}$ is true, $s$ would use a weak reference $c_\mathtt{weak}^{(s)}$ as $c_{new}^{(s)}$.

\subsection{The desired self-stabilization procedures}
\begin{figure}[htbp]
\centering
\subfloat[]{\label{fig:exchanging1}\includegraphics[width=3.3in]{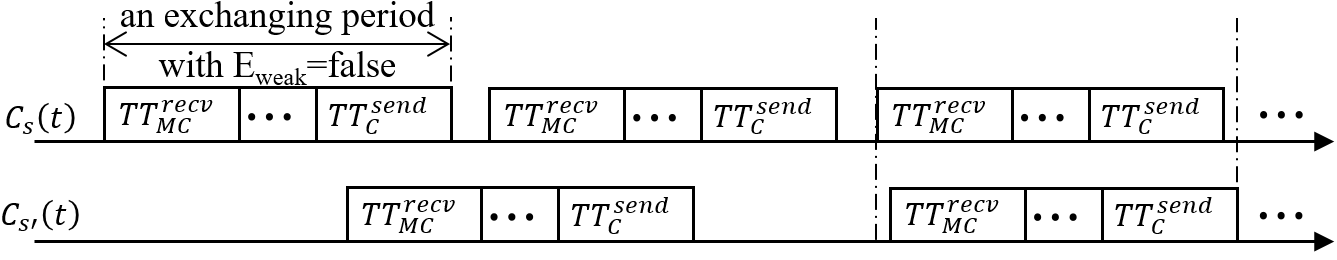}}\hfill
\subfloat[]{\label{fig:exchanging2}\includegraphics[width=3.3in]{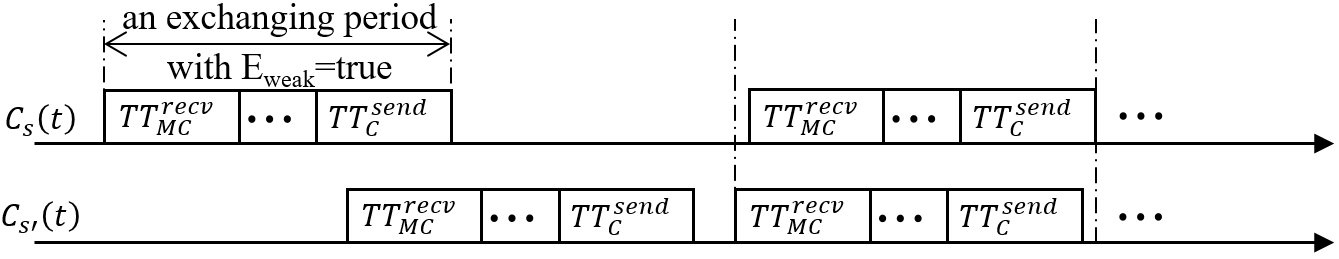}}\hfill
\caption{The desired synchronization procedures.}
\label{fig:exchanging}
\end{figure}

To put it together, we expect that some desired synchronization procedures (inspired by the hopping procedures presented in \cite{DolevWelchSelf2004}) would be performed with sufficiently high probabilities.
In Fig.~\ref{fig:exchanging}, we present two such procedures for the specific case $n_1=3$ and $f_1=1$.

By referring to the corresponding time interval of the TT-exchanging window as the exchanging period, we expect that the exchanging periods of all nodes in $Q_S$ can be eventually aligned in time.
For this, firstly, we expect that a node $s\in Q_S$ can regard itself as the \emph{grandmaster} during some exchanging periods.
For example, in the leftmost exchanging period of $s$ shown in Fig.~\ref{fig:exchanging1}, we expect that the currently tossed coin of $s$ would show $b_\mathtt{coin}^{(s)}=1$ at some instant $t_1$.
Meanwhile, we expect that every node $s'\in Q_S\setminus\{s\}$ would not be the \emph{grandmaster} during the first exchanging period of $s'$ since $t_1$ (with the $\mathtt{begin}(\mathtt{TT}_{MC}^{recv})$ tick being no earlier than $t_1$).
With this, as is shown in Fig.~\ref{fig:exchanging1}, if $\mathtt{E}_\mathtt{stb}^{(s')}$ is not true during this exchanging period of $s'$, we expect that $s'$ can be synchronized by $s$ with calling the $\mathtt{RFT}$ function in $s'$.
Otherwise, if $\mathtt{E}_\mathtt{stb}^{(s')}$ is true, as $s'$ would call the $\mathtt{FTA}$ function, we require that $s$ can find $\mathtt{E}_\mathtt{weak}^{(s)}$ being true and thus can be coarsely synchronized by $s'$, as is shown in Fig.~\ref{fig:exchanging2}.
A variant of the procedure shown in Fig.~\ref{fig:exchanging2} is that when $s$ finds $\mathtt{E}_\mathtt{weak}^{(s)}$ being true and thus adjusts $C_s$ with calling the $\mathtt{FTA}$ function, the adversary can make $\mathtt{E}_\mathtt{stb}^{(s')}$ being false in $s'$.
In this situation, for the specific case $n_1=3$, we expect that $s'$ chooses a value near that of $s$.
For the general case $n_1>2f_1$, the $\mathtt{RFT}$ function can also be extended to allow $c_{new}^{(s)}$ and $c_{new}^{(s')}$ to be sufficiently near.

Then, in all desired procedures, after the first exchanging period of $s'$ since $t_1$, we expect $C_s$ and $C_{s'}$ to be at least coarsely synchronized.
With this, during the following several exchanging periods, we expect that $s$ and $s'$ would all use the result of the $\mathtt{FTA}$ function, and thus the synchronous approximate agreement \citep{Dolev1986Approximate} would be simulated in at least $n_1-f_1$ MWS nodes.
So we expect that every node in $Q_S$ would find $\mathtt{E}_\mathtt{stb}$ being always true at the end of the desired procedure, and thus the system would be stabilized since then.

\section{Algorithm and analysis}
\label{sec:algo}
\subsection{The SS-BFT-CS algorithm}
With these strategies, we provide the algorithm $\mathtt{SSBCS}$ in Fig.~\ref{fig:advanced_sync} in solving the two dimensional SS-BFT-CS problem.

\alglanguage{pseudocode}
\algrenewcommand{\algorithmiccomment}[1]{\hskip1em//#1}
\begin{figure}[htbp]
\centering
\begin{algorithmic}[1]
\Statex \textbf{for every node $i\in Q_A$:}
\Statex \textbf{\underline{always}}
    \State  run an SSS instance as a CES node $i_p$ for every $p\in V_P$;
\Statex \textbf{\underline{on receive $m_p$ during $\mathtt{TT}_{C}^{recv}$ of $i_p$}}
    \State  $m_{p,i}=m_p$;
     ~  ${h}_{p,i}=H_i(t)\oplus \delta_{tt0}$;
     ~  $\tilde{c}_{p,i}=m_{p,i}\ominus {h}_{p,i}$;
\Statex \textbf{\underline{at $\mathtt{begin}(\mathtt{TT}_{VC}^{send})$ of $i_p$}}
    \State  $c_{i}=(\tilde{c}_{1,i}\oplus H_i(t)\oplus \delta_{tt1},\dots,\tilde{c}_{n_1,i}\oplus H_i(t)\oplus \delta_{tt1})$;
    \State  $a_{i}=(a_{1,i},\dots,a_{n_1,i})$;
    ~       $m_{i}=(m_{1,i},\dots,m_{n_1,i})$;
    \State  send $(c_{i},a_{i},m_{i})$ to $p$;
\Statex \textbf{\underline{at $\mathtt{end}(\mathtt{TT}_{C}^{recv})$ of $i_p$}}
    \State  $C_{i}(t)=\mathtt{med}(\bar{c}_{1,i},\dots,\bar{c}_{n_1,i})\oplus \delta_{tt2}$;
\Statex
\Statex \textbf{for every node $s\in Q_S$:}\Comment with $s=\mathtt{MWS}(p)$
\Statex \textbf{\underline{always}}
    \State  run an SSS instance as a WS node in switch plane $p$;\label{code:ssbcs_sss_ws}
    \If {$\tau_\mathtt{idl}=\tau_{max}$ and $C_s(t)\bmod T=0$}\label{code:ssbcs_sig_con}
    \State  $\tau_\mathtt{idl}=H_s(t)\oplus T_0$;
    ~ broadcast SIG to $p$;\label{code:ssbcs_sig}
    \EndIf
    \If {$\tau_\mathtt{idl}\ominus H_s(t) >T_0 \land \tau_\mathtt{idl}\neq \tau_{max}$}
      $\tau_\mathtt{idl}=\tau_{max}$;
    \EndIf
\Statex \textbf{\underline{at $\mathtt{end}(\mathtt{TT}_{MC}^{recv})$ of $s$}}
    \If {$b_\mathtt{coin}^{(s)}=1$}\Comment toss the coin and get the head up\label{code:ssbcs_toss}
    \State  $\textit{grand\_life}=g_0$;
    \EndIf
    \If {$\textit{grand\_life}>0$} \Comment is a grandmaster\label{code:ssbcs_life}
    \State $\textit{grand\_life}=\textit{grand\_life}-1$;
    \If {$b_\mathtt{coin}^{(s)}=0$ or $\mathtt{E}_\mathtt{stb}^{(s)}$}
     $c_{new}^{(s)}=\mathtt{FTA}(\mathbf{C}^{(s)})$;
    \ElsIf {$\mathtt{E}_\mathtt{weak}^{(s)}$}
     $c_{new}^{(s)}=c_\mathtt{weak}^{(s)}$;
    \Else
    ~ $c_{new}^{(s)}=C_{s}(t)\oplus \delta_{tt3}$;
    \EndIf
    \Else \Comment is not a grandmaster
    \If {$\mathtt{E}_\mathtt{stb}^{(s)}$}
     $c_{new}^{(s)}=\mathtt{FTA}(\mathbf{C}^{(s)})$;
    \Else
    ~ $c_{new}^{(s)}=\mathtt{RFT}(\mathbf{C}^{(s)},H_{s}(t)\oplus \delta_{tt3}\oplus \tilde{c}_{old}^{(s)})$;
    \EndIf
    \EndIf
\Statex \textbf{\underline{at $\mathtt{begin}(\mathtt{TT}_{C}^{send})$ of $s$}}
    \State  $c_{new}^{(s)}=c_{new}^{(s)}$;
    ~  distribute $m_{p}=c_{new}^{(s)}$ to $p$;
\Statex \textbf{\underline{at $\mathtt{end}(\mathtt{TT}_{C}^{send})$ of $s$}}
    \State  $\tilde{c}_{old}^{(s)}=C_{s}(t)\ominus H_{s}(t)$;
    ~  $C_{s}(t)=c_{new}^{(s)}$;
    ~  $\tau_\mathtt{idl}=\tau_{max}$;
\end{algorithmic}
\caption{Algorithm $\mathtt{SSBCS}$.}\label{fig:advanced_sync}
\end{figure}

In Fig.~\ref{fig:advanced_sync}, the algorithm is provided for the MES and MWS nodes.
As the WS nodes in $Q_{AP}\setminus Q_S$ need only to run line~\ref{code:ssbcs_sss_ws} (of the $\mathtt{SSBCS}$ algorithm, the same below) and can be trivially synchronized by the MWS node in a nonfaulty switch plane, they are ignored in the core algorithm.
With the static TT-schedules, we can configure the time parameters as $\delta_{tt0}=\mathtt{end}(\mathtt{TT}_{C}^{send})-\mathtt{begin}(\mathtt{TT}_{C}^{recv})$, $\delta_{tt1}=\mathtt{end}(\mathtt{TT}_{C}^{send})-\mathtt{begin}(\mathtt{TT}_{VC}^{send})$, $\delta_{tt2}=\mathtt{end}(\mathtt{TT}_{C}^{recv})-\mathtt{end}(\mathtt{TT}_{C}^{send})$, and $\delta_{tt3}=\mathtt{end}(\mathtt{TT}_{C}^{send})-\mathtt{end}(\mathtt{TT}_{MC}^{recv})$.
Also, the nominal \emph{SIG} cycle $T$ of an MWS node can be configured as the maximal round cycle $T_0$ (which is bounded by the time parameter $\tau_{\Phi}^{ES}$ in \cite{YuCOTS2021}) plus the bounded error $\epsilon_2$.
For randomization, the coin tossed in executing line~\ref{code:ssbcs_toss} should get the head up with the fixed probability $q_0$.
And the parameter $g_0$ can be configured as $a_0+k_0$ with $k_0=\lceil \log_{c_0}(2\epsilon_2+8\rho (1+\rho)T)/\epsilon_0\rceil$, $c_0=\lfloor (n_0-2f_0-1)/f_0 \rfloor+1$ (see \citep{Dolev1986Approximate}).
For simplicity, here we assume the lines in the algorithm are atomically executed, and the nonfaulty MES and MWS nodes can adjust their $C$ clocks at the end of the TT-slot $\mathtt{TT}_{C}^{send}$.
In practice, the $C$ clocks can also be adjusted at some other scheduled ticks after $\mathtt{end}(\mathtt{TT}_{C}^{send})$ with extra static and dynamic compensations.

Practically, although the $\mathtt{RFT}$ function can be generally extended, the $\mathtt{SSBCS}$ algorithm is best for the specific case $n_1=3$ and $f_1=1$, since $n_1$ should be sufficiently small in considering fast stabilization.
For the larger $n_1$ and $f_1$, multi-layer solutions can be built by employing the $3$-switch-plane solution as a basic building block like that in \citep{Lenzen2019AlmostasEasyas}.
As is limited here, we only provide the basic $3$-switch-plane solution.

\subsection{Analysis}
Firstly, we investigate the desired \emph{resynchronization point}.
\begin{definition}
\label{def_resynchronization}
$t$ is a resynchronization point iff line~\ref{code:ssbcs_toss} is executed in some $s\in Q_S$ at $t$ with $b_\mathtt{coin}^{(s)}=1$ and there is some $s'\in Q_S\setminus\{s\}$ with $\textit{grand\_life}=0$ during the first exchanging period since $t$.
\end{definition}

\begin{lemma}
\label{lemma_a_point}
During any time interval $[t_1,t_1+T_{max}]$ with $t_1\geqslant t_\mathtt{c}$ and $T_{max}=2T(1+\rho)$, with at least a probability $2q_0(1-q_0)^{g_0}$ that there is a resynchronization point $t\in [t_1,t_1+T_{max}]$.
\end{lemma}
\begin{IEEEproof}
(sketch)~Firstly, for every node $s\in Q_S$, as the condition checked in line~\ref{code:ssbcs_sig_con} cannot always be false in every $T+T_0$ ticks, at least one \emph{SIG} would be sent in $s$ in every $T+T_0$ ticks.
So line~\ref{code:ssbcs_toss} would be executed during every time interval $[t_1,t_1+(1+\rho)(T+T_0)]$.
So by tossing the coin, with at least a probability $q_0$ that $b_\mathtt{coin}^{(s)}=1$ when line~\ref{code:ssbcs_toss} is executed at some $t\in [t_1,t_1+T_{max}]$.
And for $t$ being a resynchronization point, it only requires that some $s'\in Q_S\setminus\{s\}$ would not get $b_\mathtt{coin}^{(s')}=1$ during the first exchanging period since $t$ and the previous $g_0-1$ exchanging periods of $s'$.
So as $|Q_S|\geqslant 2$, the overall probability is no less than $2q_0(1-q_0)^{g_0}$.
\end{IEEEproof}

With this, we show the following property of $\mathcal{S}$.
As is limited here, the proof is sketched with relaxed parameter values.
A full discussion of the exact bounds of these values is out of the range of the paper.
\begin{theorem}
\label{theorem_synchronized}
For any $t_1\geqslant t_c$, with a probability $q_1\geqslant q_0(1-q_0)^{2g_0}p_0^{g_0}(1-p_0)/2$ that $\mathcal{S}$ would be $(\epsilon_0,\rho)$-synchronized since some $t_2\in[t_1,t_1+(g_0+1)T_{max}]$ with $\epsilon_0=3(1+\rho)d_{max}$.
\end{theorem}
\begin{IEEEproof}
(sketch)~Assume $t_1$ is a resynchronization point.
As $b_\mathtt{coin}^{(s)}(t_1)=1$, $s$ would remain its $C$ clock being unadjusted unless $\mathtt{E}_\mathtt{weak}^{(s)}$ is satisfied.
If $\mathtt{E}_\mathtt{weak}^{(s)}$ is satisfied, as $n_1=3$, only two cases need to be considered.
In the first case, $s$ is in $P'$ of (\ref{eq:condition_weak}) and thus the adjustment of $C_s$ is no more than $\epsilon_2$.
In the second case, $s$ is not in $P'$ and thus all nodes in $Q_S\setminus \{s\}$ are in $P'$.
In this case, $C_s$ would be adjusted to a value with no more than $\epsilon_2$ difference to that of all other nodes in $Q_S$.
Now for anther node $s'\in Q_S\setminus \{s\}$, as $\mathtt{E}_\mathtt{weak}^{(s)}$ is satisfied in $s$, there are also two cases.
In the first case, $\mathtt{E}_\mathtt{stb}^{(s')}$ is true and thus the difference between $c_\mathtt{weak}^{(s)}$ and $c_{new}^{(s')}$ would be no more than $\epsilon_2+2\rho T_{max}$ when $s'$ adjusts $C_{s'}$ at $t_2$ during its first exchanging period since $t_1$.
With this we have $\mathring{d}(C_s(t_2),C_{s'}(t_2))\leqslant 2\epsilon_2+4\rho T_{max}$.
In the second case, $\mathtt{E}_\mathtt{stb}^{(s')}$ is not true and thus $\mathtt{RFT}$ would be called in $s'$.
So if $s$ is not in $P'$, with a probability $(1-p_0)/4$ $s'$ would choose a value in $\{c_{s'},c_{s'}^{pre}\}$ with no more than $\epsilon_2+2\rho T_{max}$ difference to $c_{new}^{(s)}$ and thus $\mathring{d}(C_s(t_2),C_{s'}(t_2))\leqslant 2\epsilon_2+4\rho T_{max}$ holds.
And if $s$ is in $P'$, there is also a probability $(1-p_0)/4$ that $s'$ would choose the value of $s$ and thus $\mathring{d}(C_s(t_2),C_{s'}(t_2))\leqslant 2\epsilon_2+4\rho T_{max}$ would be true.

Otherwise, if $\mathtt{E}_\mathtt{weak}^{(s)}$ is not satisfied in $s$, as the instant corresponding to $\mathtt{begin}(\mathtt{TT}_{MC}^{recv})$ of $s'$ is no earlier than $t_1$, both $\mathtt{E}_\mathtt{stb}^{(s)}$ and $\mathtt{E}_\mathtt{stb}^{(s')}$ would be false.
So $s$ would remain its $C$ clock being unadjusted during the current exchanging period, and there is also a probability $(1-p_0)/4$ that $s'$ would choose the value of $s$ in executing the $\mathtt{RFT}$ function.

So, at the end of the first exchanging period of $s'$ since $t$, there is at least a probability $(1-p_0)/4$ that $\mathring{d}(C_s(t_2),C_{s'}(t_2))\leqslant 2\epsilon_2+4\rho T_{max}$.
Then, during the following $g_0-a_0$ rounds, with a probability $((1-q_0)p_0)^{g_0-a_0}$ the approximate agreement would be simulated in $s$ and $s'$.
So $\mathring{d}(C_s(t),C_{s'}(t))\leqslant \epsilon_0$ holds at the end of the $g_0-a_0$ rounds.
Similarly, with another probability $((1-q_0)p_0)^{a_0}$ that $\mathtt{E}_\mathtt{stb}^{(s'')}$ would be true in every node $s''\in Q_S$ at the end of the $g_0$ rounds and thus the system is $(\epsilon_0,\rho)$-synchronized since then.
And with Lemma\ref{lemma_a_point}, the overall probability of this is no less than $2q_0(1-q_0)^{g_0}((1-q_0)p_0)^{g_0}(1-p_0)/4$.
\end{IEEEproof}

\subsection{Discussion}
With the proof of Theorem~\ref{theorem_synchronized}, the expectation of $\Delta_\mathtt{stb}$ (denoted as $\Delta_\mathtt{stb}^{(exp)}$) can be approximately $T_{max}/q_1+g_0 T_{max}$.
To maximize $q_1$, we can set $q_0=1/(2g_0+1)$ and $p_0=1-1/(g_0+1)$ and get $q_1> 1/(2e^2(2g_0+1)(g_0+1))$, where $e< 2.72$ is the Euler's number.
So $\Delta_\mathtt{stb}^{(exp)}$ is mainly depends on $g_0=a_0+k_0$.
For a smaller $a_0$, as $c_0=\lfloor (n_0-2f_0-1)/f_0 \rfloor+1$, we can increase $n_0$.
Namely, $\Delta_\mathtt{stb}^{(exp)}$ can be lowered by deploying more MES nodes in the network.
And when $n_0$ is sufficiently large, we would get $g_0=a_0+1$, which would make $\Delta_\mathtt{stb}^{(exp)}$ being independent of $f_0$.
Then, as $T_{max}\approx 2T$, we can set $a_0\leqslant 3$.
With these, we can approximately get $\Delta_\mathtt{stb}^{(exp)}\leqslant (2/q_1+4)T$ with $q_1>1/(90e^2)$.
As the nominal \emph{SIG} cycle $T$ can be at the order of milliseconds or even sub-millisecond, $\Delta_\mathtt{stb}^{(exp)}$ can be at the order of a second or better.
Also, as $(1-q_1)^{10^4}< 3\times 10^{-7}$ with $q_1>1/(90e^2)$, the probability that $\mathcal{S}$ is not $(\epsilon_0,\rho)$-synchronized before $(10^4+4) T_{max}$ is less than $3\times 10^{-7}$.
So by taking $T_{max}$ as approximately $1$ millisecond, $\mathcal{S}$ can be synchronized in less than $10$ seconds with a very high probability.

In comparing with existing solutions, firstly, the solution provided here is for tolerating Byzantine faults in both the terminal components and the communication components.
So in considering Byzantine-resilience, this solution is inherently better than TTP, TTEthernet, and the basic WALDEN solution where the failure-modes of the communication components are strictly restricted.
Also, as only $n_1>2f_1$ is required, this solution is better than that of synchronizing the switches and then distributing their clocks to the terminals, since that would require $n_1>3f_1$.
In considering the stabilization time, as we do not rely on any BA routine nor fully connected network, the stabilization time is independent of $f_0$, and thus the solution provides better scalability of $n_0$ than that of the classical ones \citep{DolevPulseBoundedDelay2007,Dolev2014PulseGeneration,Lenzen2019AlmostasEasyas}.
Also, as the communication and computation required in the $\mathtt{SSBCS}$ algorithm can be optimized by static TDMA schedules, the required system resources can be minimized in each switch plane.
With this, the SS-BFT-CS solution provided in the advanced WALDEN networks can be more efficient than the traditional SS-BFT-CS solutions where the communication might be asynchronous when the system is not synchronized.

\section{Conclusion}
\label{sec:con}
In this paper, we have investigated the two-dimensional SS-BFT-CS problem and solved this problem with the WALDEN networks.
The WALDEN networks are specially investigated as they are composed of common COTS Ethernet components.
Now with the provided SS-BFT-CS solution, more reliable distributed CS systems can be built with common COTS Ethernet components.
This would help in developing distributed hard-real-time systems with stringent time, resources, and safety requirements.

Firstly, to make the SS-BFT-CS solution decoupled with the underlying networks, we utilize the TT stages provided by the underlying networks to establish upper-layer TT-exchanging windows.
As we do not rely on any concrete realization of the TT-exchanging windows, the multi-plane SS-BFT-CS solution is independent of the realization of the underlying basic CS solution.
Then, to align the TT stages in all nonfaulty switch planes, we have developed a randomized synchronization scheme to synchronize the $C$ clocks of the nonfaulty MES and MWS nodes.
We have shown that the provided SS-BFT-CS algorithm can reach stabilization in an expected small duration independent of $f_0$ with $n_1=3$, $f_1=1$, and a sufficiently large $n_0$.
In applying to WALDEN networks, the system can be synchronized in one or several seconds with high probabilities.

Despite the merits, the provided solutions can be improved in several ways.
Firstly, to tolerate more faulty switch planes, multi-layer solutions can be further explored.
Secondly, for deterministic stabilization, the basic method can be further integrated with BA-based SS-BFT-CS solutions.
Also, for better synchronization precision, the basic method can be integrated with high precision protocols like PTP \citep{ieee1588v2}.

\bibliographystyle{IEEEtran}
\bibliography{IEEEabrv,TDWALDEN}

\end{document}